\documentclass[dvips]{article}
\usepackage{cite}
\usepackage{icrctc07}

\title{Hard particle spectra from parallel shocks due to 
turbulence transmission}

\shorttitle{Turbulence transmission in shocks}

\authors{Joni Tammi}

\shortauthors{Joni Tammi}

\afiliations{UCD School of Mathematical Sciences, 
  University College Dublin, Dublin 4, Ireland}
\email{joni.tammi@iki.fi}

\abstract{If taken into account, the transmission of the
  particle-scattering turbulence --in addition to just the particles--
  through the shock front can change the effective compression ratio
  felt by the accelerating particles significantly from the
  compression of the underlying plasma.  This can lead to
  significantly harder energy spectra than what are traditionally
  predicted assuming frozen-in turbulence. I consider the
  applicability and limitations of turbulence transmission scenario in
  parallel shock waves of different thickness, its consequences in AGN
  and microquasar environments, and discuss the possible effects to
  the spectrum of the accelerated particles. }

\begin{document}
\maketitle

%
%
\section{Introduction}

The first-order Fermi acceleration process in shock waves is often
considered as the main mechanism responsible for the nonthermal
electron populations assumed to produce the observed radiation in many
astrophysical sources. In the basic theory partices gain energy
scattering elastically off magnetic turbulence --e.g., Alfv\'en
waves-- frozen-in to
 the converging flow. The process is known to
accelerate charged particles to power-law energy distributions $N
\propto E^{-\sigma}$, with spectral index $\sigma$ having value $\sim
2$ for nonrelativistic and $\sim 2.2$ for relativistic shocks, fitting
many observations.

For the simplest nonrelativistic (unmodified) step shocks the spectral
index of the accelerated particles is known to depend mainly on the
compression ratio of the plasma, 
\( r \simeq r_{\rm flow} = V_1 / V_2 \) ,
as
\( \sigma = \frac{r + 2}{r - 1}, \)
where $V_1$ and $V_2$ are the shock-frame speeds of the far up- and
downstream flows, respectively. The compression ratio itself depends
on the hydrodynamics of the flow, and has values ranging from $4$
(nonrelativistic) to $3$ (for $V_1 \to c$). In modified thicker shocks
the effect of broadened speed profile leads to decreased acceleration
efficiency and steeper spectra.\cite{SchneiderKirk1989,VV2003ICRC}

Although many observations are in accordance with $\sigma$ of $2$ or
$2.2$, some sources seem to require significantly smaller indices
beyond the limits of the traditional first-order acceleration
theory. Although there are other plausible mechanisms cabable of
producing $\sigma < 2$, omitting the assumption of frozen-in
turbulence might be sufficient even for the first-order process alone
to allow for hard spectra.  Namely, by including the dynamics of the
particle-scattering waves --and the effects the shock has on them-- in
the analysis, one arrives to a situation where the \emph{effective}
compression ratio felt by the scattering centres can be significantly
higher than that of the underlying flow, $r_{\rm flow}$.%
\cite{VainioSchlickeiser1998,VVS2003,VVS2005,VV2005AA,TV2006,Tammi2006PhD}

In this paper I review recent studies done so far for parallel shocks,
discuss the physical requirements for the scenario, and illustrate the
effect method in the case of the microquasar Cygnus X-3.

%
%
\vspace{-0.5em}
\section{Wave transmission}
\vspace{-0.2em}


In very thin astrophysical plasmas particle--particle collisions are
extremely rare and the particles only ``see'' the magnetic
turbulence. If this turbulence is frozen-in to the plasma, then the
scattering-centre speed is simply that of the underlying plasma
flow. But, as has been known for long \cite{Bell1978a}, if the waves
themselves have significant speeds with respect to the flow, then also
the scattering-centre speed changes. For turbulence consisting of
Alfv\'en waves propagating in the plasma with Alfv\'en speed $V'_{\rm
A}$ either parallel (``forward'') or antiparallel (``backward'') to
the direction of the flow with respect to the shock, the wave speed in
the shock rest frame is simply the flow speed $V(x)$ in the shock
frame plus (or minus) the local Alfv\'en speed in the plasma
frame. When there are both wavemodes present, it becomes useful to
define the \emph{normalised cross-helicity} of the waves as
\[ H_c = \frac{ I^+ - I^- }{ I^+ + I^- } \in [-1, +1], \]
%
where \( I^\pm \equiv I^\pm(x,k) \propto k^{-q} \) (for $k > k_0$)
are the wave spectra of the forward ($+$) and the backward ($-$) waves
for a given wavenumber at a given location. The ``mean scattering-centre
speed'' is then
\[ V_{\rm sc} = \frac{V + H_c V'_{\rm A}}{1 + H_c V V'_{\rm A}}. \]
%

Furthermore, we introduce the the Lorentz factor \( \Gamma =
1/\sqrt{1-V^2/c^2} \) and the \emph{quasi-Newtonian Alfv\'enic Mach
number},
\[ M \equiv \frac{ V_{1}\Gamma_{1} }{ V'_{{\rm A1}}\Gamma_{{\rm A1}} }, \]
where the proper Alfv\'en speed 
%
\[ V'_{{\rm A1}}\Gamma_{{\rm A1}} = \frac{B}{\sqrt{4\pi \mu n}}, \] 
and the specific enthalpy \( \mu=(\rho + P)/n \) depend on the total
energy density $\rho$, number density $n$, and the pressure, $P$, of
the gas.

In a super-Alfv\'enic shock the speed of the flow with respect to the
shock is always larger than the local wave speed. In other words, as
seen in the shock rest frame, both the forward and backward moving
waves propagate toward the shock in the upstream and away from it in
the downstream; there are no waves crossing the shock from downstream
to upstream. This allows us to define the \emph{critical Mach number}
$M_{{\rm c}}\equiv\sqrt{r}$, above which the aforementioned conditions
are fulfilled.~\cite{VVS2003}

In order to calculate the scattering-centre speeds on both sides of
the shock one has to solve how the shock crossing affects the
wavelenghts and amplitudes of the waves. So far this has been done
separately for the step shock approximation (i.e., for waves much
longer than the shock structure)
\cite{VainioSchlickeiser1998,VVS2003,VVS2005} and for thick modified
shocks (or for waves sufficiently short, seeing the transition of flow
parametres as smooth).~\cite{VV2005AA,TV2006}

%
\subsection{Wave field behind the shock}

Details of the transmission process depend on the wavelength of the
crossing wave; if the waves are long compared to the shock transition,
the part of the waves are simply transmitted through the shock while
part is reflected (i.e., ``$+$'' waves become ``$-$'' waves and vice
versa) and regardless of the upstream wave composition there will be
both wave modes present in the behind the
shock.~\cite{VainioSchlickeiser1998,VVS2003} Different waves are also
amplified differently leading to situation where, regardless of the
upstream wave field, the waves immediately behind a step shock are
flowing predominantly antiparallel to the flow and ``following'' the
shock.

If the shock transition is wide enough to allow for the waves to see
it as a smooth change from upstream to downstream plasma parametres,
there will in general be no wave reflection case (see, however, e.g.
\cite{Laitinen2005}), and an all-forward upstream will result into
all-forward downstream and likewise for the backward waves (i.e., \(
H_{c1} = \pm 1 \Rightarrow H_{c2} \pm 1\)).~\cite{VV2005AA,TV2006}
This leads to a qualitative difference compared to the step-shock
case as now, depending on the upstream wave field and the shock
speed, there are also cases where the downstream cross-helicity (or
the ``average wave direction'') is positive, i.e., most of the waves
are propagating forward (see Fig.~1 in \cite{TV2006}).

%
\subsection{Increased compression ratio}

\begin{figure}
\includegraphics[width=\linewidth]{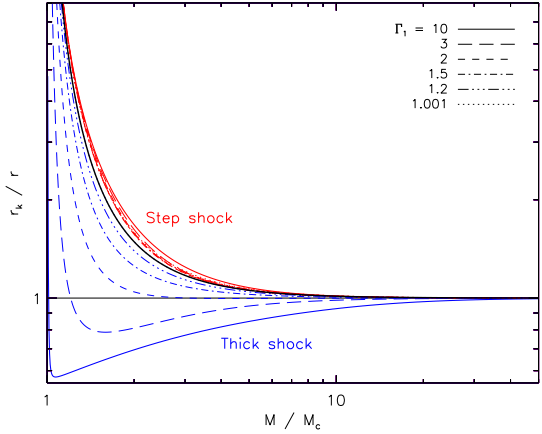}
\caption{\small Scattering-centre compression ratio $r_k$ as a
  function of Alfv\'enic Mach number $M$ for various shock speeds in
  the case of upstram turbulence corresponding to $H_{c1} =0$ and
  $q=5/3$. The thick black line separates the step and thick shock
  cases.}
\end{figure}

From the wave and flow speeds and the cross-helicity on both sides of
the shock one can calculate the compression ratio of the scattering
centres, 
\( r_k \equiv V_{k1} / V_{k2}. \)
This is the compression felt by the particles crossing the
shock, and it can differ significantly from the compression ratio 
of the flow $r$. For $M \to \infty$ the effective wave speed $V_k
\to V$ and we have usual the frozen-in turbulence, but as $M \to M_c$
(or, as $V'_{\rm A}$ becomes non-negligible with respect to $V$), the
scattering-centre compression ratio starts to differ from that of the
flow.

Figures 1 and 2 show $r_k / r$ as a function of the
Alfv\'enic Mach number (normalised to the critical value $M_c$) for
different shock speeds; Fig.~1 corresponds to the case with
Kolmogorov-type turbulence pre-existing in the upstream ($H_{c1}=0,
q=5/3$), and Fig.~2 has all waves initially antiparallel to the flow,
as if self-generated by high-energy cosmic rays streaming ahead of the
shock ($H_{c1}=-1, q=2$). In Fig.~1 one can also see the case for
thick relativistic shocks (or very short waves), where the effective
compression ratio can also be decreased. In most cases the compression
ratio is, however, increased for magnetic field strengths sufficient
to allow for non-negligible wave speeds.

\begin{figure}
\includegraphics[width=\linewidth]{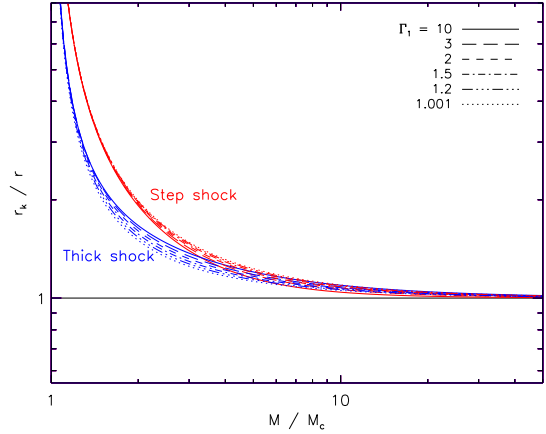}
\caption{\small Same as Fig.~1 but for $H_{c1} = -1$ and $q=2$.}
\end{figure}

%
%
\section{Example: Cygnus X-3 flare}

A recent study of a microquasar Cygnus X-3 flare suggests the presence
of electron population with a high-energy power-law distribution with
energy spectral index $\sigma \approx 1.77$.\cite{LindforsEtAl2007}
Let us apply the wave transmission scenario to see what would be
required for the physical properties if such a spectrum would be
produced by a simple strong step shock moving with the deduced speed
$V_1 = 0.63$ c into a low-density plasma with pre-existing Kolmogorov
type turbulence. In the following we set $c=1$, and follow the
transmission analysis presented in detail in \cite{VVS2003,VVS2005}.

Inversing the equation for $\sigma(r)$ to $r(\sigma)$ we can see that
if the $\sigma=1.77$ is assumed to be due to first-order process,
compression ratio of $r_{k}\approx4.90$ would be needed.\footnote{Note
that the equation for spectral index is not accurate for other than
the simplest nonrelativistic shocks -- for faster speeds it
overestimates the required compression. This value is still
illustrative as it sets even stricter limits for the magnetic
field. Furthermore, the nonrelativistic formula for diffusive
acceleration has been observed to work rather well even in
relativistic shocks if increased compression is
present.\cite{VV2005AA}} Describing the plasma as a dissipation-free
ideal gas, the compression ratio of the flow itself is only $r_{\rm
flow}\approx3.78$, so the scattering-centre compression ratio would
have to be $r_{k}/r_{\rm flow}\approx1.24$ times that of the flow.
\begin{figure}
\includegraphics[width=\linewidth]{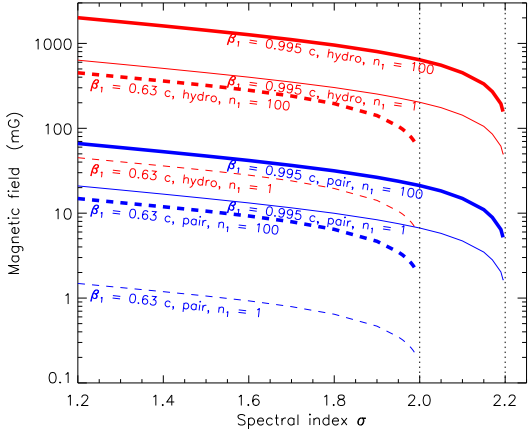}
\caption{\small Resulting particle spectral index $\sigma$ from
  parallel step shocks when the upstream turbulence ($q=5/3,
  H_{c1}=0$) is transmitted through a step shock and the resulting
  effective compression ratio is used for acceleration. Solid lines
  are for a relativistic shock moving with speed $V_1 = 0.995$ c,
  dashed lines correspond to $V_1 = 0.63$ c (Cyg X-3); upstream number
  densities are $n=1\,{\rm cm^{-3}}$ and $100\,{\rm cm^{-3}}$ (thin and
  thick lines, respectively) The two vertical dotted lines mark the
  limit for frozen-in turbulence.}
\end{figure}

From Fig.~1 we can see that this kind of increased compression in the
case of a mildly relativistic step shock could follow if the ratio of
Alfv\'enic Mach number to the critical Mach number $M_c = \sqrt{r_{\rm
    flow}}$ is
\( M/M_c \approx 2.74 \, \Rightarrow \, M \approx 2.74 \sqrt{r_{\rm flow}} 
\approx 5.33. \)
Now, combining the equations for the Alfv\'enic proper speed and the
Mach number, we get the magnetic field: 
\( B=\Gamma_{1}V_{1}\sqrt{4\pi\mu n}/M. \) 

For a strong shock the pressure of the upstream gas is negligible to
its rest energy, so the specific enthalpy becomes $\mu = m$, where the
mass $m$ depends on the composition of the plasma: $m=m_{{\rm
p}}+m_{{\rm e}}$ for hydrogen plasma, and $m=2m_{{\rm e}}$ for
electron-positron pair plasma. As neither the composition nor the
number density is well known, we use $n=1\,{\rm cm^{-3}}$.  Substituting
the values in to the equation of the magnetic field, we get
\( B_{{\rm H}}          \approx 21   \,{\rm mG} \) and 
\( B_{{\rm {\rm pair}}} \approx 0.69 \,{\rm mG} \).
Both agree well with the upper limits \textbf{$B<0.15\,{\rm G}$}
deduced from observations of \cite{MillerJonesEtAl2004}, so with the
aforementioned assumptions the increased compression ratio due to
turbulence transmission could play a role in explaining the hard
spectrum. It is interesting to note that if the value $100\,{\rm
cm^{-3}}$ \cite{MillerJonesEtAl2004} is used for the upstream number
density, the fields become
\( B_{{\rm H}}         \approx 0.21 \,{\rm G} \) and 
\( B_{{\rm {\rm pair}}} \approx 6.9 \,{\rm mG} \), 
rising the $B_{{\rm H}}$ beyond the upper limit. More detailed
knowledge of the plasma composition and the magnetic field could yield
a tool for testing this model for both the improving and rebuttal
purposes.

The relation between the spectral index and the magnetic field is
shown in Fig.~3 for the Cyg X-3 case and additionally for a fully
relativistic shock with $\Gamma_1 \approx 10$, and for densities $1$
and $100\,{\rm cm^{-3}}$.

%

%
%
\vspace{-0.2em}
\section{Conclusions}
\vspace{-0.3em}

I have discussed particle acceleration by the first-order Fermi
acceleration mechanism leaving out the simplifying assumption of
frozen-in turbulence. I have described the basic results of Alfv\'en
wave transmission analysis
\cite{VainioSchlickeiser1998,VVS2003,VVS2005,VV2005AA,TV2006,Tammi2006PhD},
and the requirements for it to have significant changes in the
accelerated particle spectrum, with emphasis on the possibility of
having a natural way for parallel shocks to produce spectra steeper
than the ``universal'' $\sigma \approx 2.2$ value. I have also
demonstrated that the physical properties required can be close
to those observed, suggesting that this kind of
mechanism could indeed be active in certain sources.

I also stress that the mechanisms discussed rely on various
assumptions. First of all, the current theory only deals with parallel
shocks. It also neglects the particle--wave, wave--shock and
wave--wave interactions, and assumes the turbulence to consist of
small-amplitude Alfv\'en waves. Furthermore, the short and long
wavelength waves (and particle scattering off them) are treated
separately, although it is probable for any real astrophysical shock
to affect waves of all lengths. Finally, the second-order Fermi
acceleration, very likely to affect the produced spectra, is not
considered here, as its connection to the transmission analysis has
been studied elsewhere.\cite{VV2005ApJ}

%
%
\bibliography{icrc0530}
\bibliographystyle{prsty}

\end{document}